\documentclass[aps,prl,preprint,superscriptaddress]{revtex4-1}

\usepackage{graphicx}
\usepackage{dcolumn}
\usepackage{bm}
\usepackage{footnote}

\begin{document}


\title{Beryllium Polyhydride Be$_{4}$H$_{8}$(H$_{2}$)$_{2}$ Synthesized at HP/HT}


\author{Takahiro Matsuoka}
\email[]{TKMatsuoka08@gmail.com/tmatsuok@utk.edu}
\affiliation{Department of Electrical, Electronic and Computer Engineering, Faculty of Engineering, Gifu University, Gifu 501-1193, Japan}
\affiliation{Material Science and Engineering, Joint Institute for Advanced Materials (JIAM), The University of Tennessee, Knoxville, TN 37996, USA}

\author{Hiroshi Fujihisa}
\email[]{hiroshi.fujihisa@aist.go.jp}
\affiliation{National Institute of Advanced Industrial Science and Technology (AIST), Ibaraki 305-8565, Japan}

\author{Takahiro Ishikawa}
\affiliation{The Elements Strategy Initiative Center for Magnetic Materials, National Institute for Material Science, Ibaraki 305-0047, Japan}

\author{Takaya Nakagawa}
\affiliation{Graduate School of Engineering, Gifu University, Gifu 501-1193, Japan}

\author{Keiji Kuno}
\affiliation{Environmental and Renewable Energy Systems Division, Gifu University, Gifu 501-1193, Japan}

\author{Naohisa Hirao}
\affiliation{Japan Synchrotron Radiation Research Institute (JASRI), Hyogo 679-5198, Japan}

\author{Yasuo Ohishi}
\affiliation{Japan Synchrotron Radiation Research Institute (JASRI), Hyogo 679-5198, Japan}

\author{Katsuya Shimizu}
\affiliation{Center for Science and Technology under Extreme Conditions, Graduate School of Engineering Science, Osaka University, Osaka 560-8531, Japan}

\author{Shigeo Sasaki}
\affiliation{Department of Electrical, Electronic and Computer Engineering, Faculty of Engineering, Gifu University, Gifu 501-1193, Japan}

\date{\today}

\begin{abstract}
We report the XRD and Raman scattering measurements in combination with DFT calculations that reveal the formation of beryllium polyhydride Be$_{4}$H$_{8}$(H$_{2}$)$_{2}$ by the laser heating of Be/H$_{2}$ mixture to above 1700 K at pressures between 5 GPa and 8 GPa. The Be$_{4}$H$_{8}$(H$_{2}$)$_{2}$ crystallizes in a \textit{P}6$_{3}$/\textit{mmc} structure and consists of the corner-sharing BeH$_{4}$ tetrahedrons and the H$_{2}$ molecules that occupy an interstitial site. The Be$_{4}$H$_{8}$(H$_{2}$)$_{2}$ is stable at least to 14 GPa on compression and stable down to 4 GPa at room temperature. Our \textit{ab-initio} calculations suggest that the Be$_{4}$H$_{8}$(H$_{2}$)$_{2}$ is a meta-stable phase of the Be-H system.
\end{abstract}


\maketitle

Since the discovery of beryllium dihydride (BeH$_{2}$) \cite{Barbaras1951}, extensive studies of physical and chemical properties have been performed. 
Although the toxicity of Be hampers the practical use, the high gravimetric H$_{2}$ density (Hwt\% = 18.3\%) is attractive for hydrogen storage. 
Besides, with only six electrons and light mass, BeH$_{2}$ has been an excellent bench to test advanced quantum molecular calculations \cite{Sampath2003a,Koukaras2016,Hrenar2005,BheemaLingam2011,Vidal2006}. 
Furthermore, significant Hwt\% makes the investigation of superconductivity appealing because a strong electron-phonon coupling and the high frequency of atomic vibration are anticipated. 
In these contexts, finding the hydrogen richer allotropes of the Be-H system is a particular interest.
However, there has been no experimental and theoretical studies that have found any stable BeH$_\textit{n}$ (\textit{n} $>$ 2).

BeH$_{2}$ has a covalent or mixed covalent/ionic Be-H bond \cite{Greenwood1984,Wang2010a}. 
The ionicity is mainly due to the charge transfer from Be 2\textit{s} to H 1\textit{s} atomic orbitals, while the hybridization of H 1\textit{s} and Be 2\textit{p} states dominates the covalency \cite{Greenwood1984}. 
In the solid-state, BeH$_{2}$ molecules share their H atoms with neighboring BeH$_{2}$ molecules and form corner-sharing BeH$_{4}$ tetrahedrons \cite{Sampath2003a,Marchenko1982}. 
The BeH$_{4}$ tetrahedrons form two types of BeH$_{2}$ allotropes at ambient pressure. 
One is an amorphous BeH$_{2}$ \cite{Brendel1978}. 
Another is a crystalline BeH$_{2}$ (called $\alpha$-BeH$_{2}$) with an \textit{Ibam} (body-centered orthorhombic) structure.
The $\alpha$-BeH$_{2}$ is synthesized by compacting amorphous BeH$_{2}$ at high temperatures \cite{Smith1988, Pepin2016}. 
The phase diagram of BeH$_2$ has been revealed up to 100 GPa at temperatures from 300 to 1500 K by heating the Be/H$_2$ mixture \cite{Pepin2016}.
P\'{e}pin \textit{et al}. have observed the synthesis of another BeH$_{2}$ (named $\beta$-BeH$_{2}$, \textit{P}4$_{1}$2$_{1}$2 structure) along with $\alpha$-BeH$_{2}$ by keeping the Be/H$_{2}$ mixture at 2.4 GPa and 550 K for several ten hours \cite{Pepin2016}. 
The $\alpha$-BeH$_{2}$ is stable up to 27 GPa, while $\beta$-BeH$_{2}$ transforms into a different phase (called $\beta$'-BeH$_{2}$) at 17 GPa \cite{Pepin2016}. 
Between 27 GPa and 72 GPa, any BeH$_{2}$ allotrope decomposes into Be and H$_{2}$ \cite{Pepin2016}. 
Be and H$_{2}$ react again and form a BeH$_{2}$ that crystallizes in a \textit{P}$\bar{3}$\textit{m}1 layered structure (1T-structure) at 72 GPa \cite{Pepin2016}. 
Any other phases have not been obtained below 1500 K.
The Raman scattering measurements on amorphous BeH$_{2}$ reported a structural transition at 80 GPa \cite{Nakano2005}. 
Theoretical calculations have predicted several structural transformations of BeH$_{2}$ from the \textit{Ibam}, including the transformation to the \textit{P}$\bar{3}$\textit{m}1 layered structure, at pressures between 0 and 400 GPa \cite{Vajeeston2004, Wang2014, Yu2014, Hooper2013}. 
However, a calculation using first-principles variable-composition evolutionary methodology has suggested that BeH$_{2}$ remains the only stable phase up to 400 GPa \cite{Yu2014}. 
An evolutionally structure search coupled with density functional theory (DFT) calculation has also predicted that any phases of BeH$_{\textit{n}}$ (\textit{n} $>$ 2) are unstable toward decomposition into H$_{2}$ and BeH$_{2}$ to at least 250 GPa \cite{Hooper2013}. 
According to the calculations, BeH$_2$ remains insulating or semiconducting below 200 GPa \cite{Vajeeston2004,Wang2014,Yu2014}. 
Above 200 GPa, \textit{Cmcm} (\textit{P} $\geq$ 200 GPa) and \textit{P}4/\textit{nmm} ( \textit{P} $\geq$ 350 GPa) phases exhibit superconducting transitions at temperatures between 20 and 90 K \cite{Wang2014, Yu2014}.

This study aims at searching for hydrogen-rich BeH$_{\textit{n}}$ (\textit{n} $>$ 2) under high-pressure and high-temperature H$_2$ atmosphere by powder X-ray diffraction (XRD), Raman spectroscopy, and DFT calculation.

In the XRD and Raman scattering measurements, we used a diamond-anvil cell (DAC) with type-Ia diamond anvils with 0.4 mm culets. 
We pre-indented 0.2 mm thick gaskets made of rhenium (Re) to 50${-}$60 $\mu$m and drilled through a 150 $\mu$m diameter hole (sample chamber) at the center of the gaskets.
We loaded a tiny piece of Be (Sigma-Aldrich, 99.9\%) in the sample chamber and filled the remaining room with fluid H$_{2}$ using a cryogenic H$_{2}$-loading system \cite{Chi2011}. 
The Be to H$_{2}$ molar ratios in three independent experiments were between 1:6 and 1:12.
We compressed the small chips of ruby (Al$_{2}$O$_{3}$, corundum) together with the sample and estimated the pressure by ruby fluorescence wavelength using a proposed calibration curve \cite{Mao1986}. 
We compressed the Be/H$_{2}$ mixture at room temperature to the pressures between 5 and 8 GPa, and then we heated it to above 1700 K by focusing IR laser ($\lambda$ = 1070 nm, SPI laser) on Be. 
We collected the thermal radiation from the sample and analyzed it in a wavelength between 600 and 800 nm to convert the spectrum to temperature following Planck's blackbody radiation law \cite{Ohishi2008}. 
The reaction between Be with H$_{2}$ was detected by monitoring the crystal structure change of Be from hcp using \textit{in-situ} XRD measurements at BL10XU/SPring-8. 
After the hcp structure of Be completely disappeared, we quenched the sample to room temperature. 
We performed Raman scattering measurements at room temperature in a backscattering geometry by a triple polychromator (JASCO NR1800) equipped with a liquid-nitrogen-cooled charge-coupled device. 
The 532 nm radiation from a solid-state laser was used for excitation. 
The focused laser spot size on the samples was about 10 $\mu$m in diameter.

Figure 1a shows the representative XRD profiles before and after the laser heating at 7.1 GPa, together with the result of Rietveld refinement for the suggested structure model discussed later. 
Note that we first compressed the Be/H$_2$ sample to 7.1 GPa and performed XRD measurement at room temperature (Fig. 1a upper panel).
Then, we conducted the laser heating and quenched the sample to room temperature.
After quenching, we measured the XRD at room temperature (Fig. 1a lower panel).
Before the laser heating, we observed Be and ruby.
After the heating, the XRD peaks of Be disappeared, and several new peaks appeared. 
All of the newly emerged peaks can be indexed with a hexagonal \textit{P}6$_{3}$/\textit{mmc} structure (\textit{a} = 4.300(1) \AA, \textit{c} = 6.997(3) \AA, \textit{V} = 112.05(6) \AA), in which Be atoms are at an atomic position (1/3, 2/3, 0.5610) $[$\textit{Z} (Be) = 4$]$.
We temporary name this phase h-BeH$_{\textit{n}}$. 
Figure 1b is the volumes per formula unit (\textit{V}$_{f.u.}$) vs. pressure.
The \textit{V}$_{f.u.}$ is calculated by dividing the unit cell volume by the number of BeH$_{2}$ units.
The \textit{V}$_{f.u.}$ of h-BeH$_{\textit{n}}$ is larger and less compressible than that of $\alpha$-BeH$_{2}$, suggesting that \textit{n} exceeds 2.

Figure 1c indicates the Raman scattering spectra measured on h-BeH$_{\textit{n}}$. 
Since h-BeH$_{\textit{n}}$ was surrounded by excess H$_{2}$, we observed the signals from both h-BeH$_{\textit{n}}$ and solid H$_{2}$. 
The Raman scattering peaks from h-BeH$_{\textit{n}}$ are easily distinguishable from H$_{2}$ using the data obtained in excess H$_{2}$.
For a control experiment, we heated Be in argon (Ar) to 2000 K at 5 GPa and confirmed that Be kept hcp structure in agreement with previous XRD experiments \cite{Lazicki2012a}.
In the Raman scattering spectra of h-BeH$_{\textit{n}}$, there is a broad peak centering at 1960 cm$^{-1}$. 
A theoretical study has calculated that the Be-H stretching modes frequencies of the BeH$_{4}$ tetrahedron in $\alpha$-BeH$_{2}$ are about 1750 (symmetric mode) and 1990 cm$^{-1}$ (anti-symmetric mode) \cite{Hantsch2003}. 
Experimentally, these modes are observed at wavenumbers between 1300 and 2000 cm$^{-1}$ for an $\alpha$ and $\beta$-BeH$_{2}$ mixture at 4 GPa, and they  shift to higher wavenumbers under further compression \cite{Pepin2016,Sampath2008}. 
The observed peak at 1960 cm$^{-1}$ at 7.1 GPa suggests the presence of BeH$_{4}$ tetrahedron in the unit cell of h-BeH$_{\textit{n}}$.

\begin{figure*}
\includegraphics[scale=0.9]{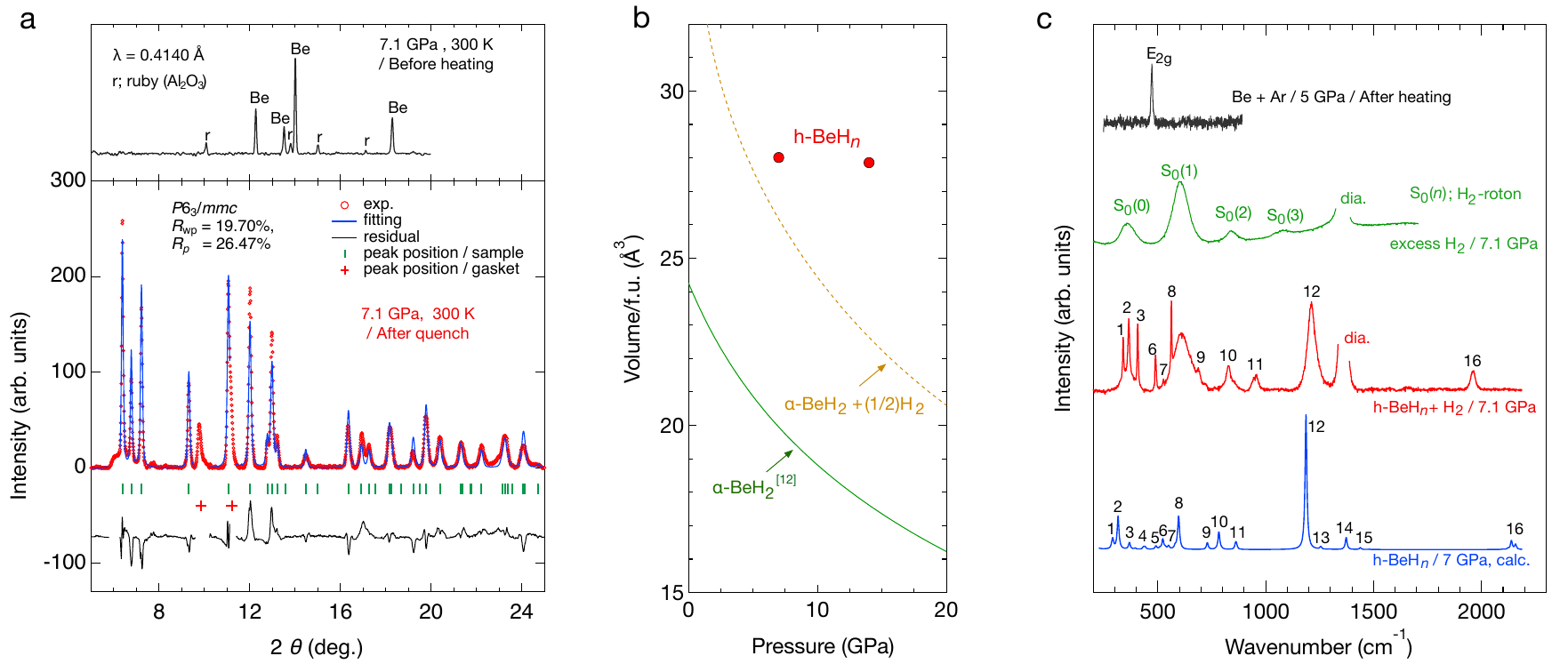}
\caption{\label{fig1}(color online) (a) XRD profiles of the Be/H$_{2}$ mixture before and after the laser heating with Rietveld fitting result for the suggested structure model. (b) The \textit{V}$_{f.u.}$ vs. \textit{P} for  h-BeH$_{\textit{n}}$ (filled circle), $\alpha$-BeH$_{2}$ (solid line), and $\alpha$-BeH$_{2}$+(1/2)H$_2$ (dashed line). The \textit{V}$_{f.u.}$ of $\alpha$-BeH$_{2}$+1/2(H$_2$) is calculated using the data shown in references \cite{Pepin2016, Loubeyre1996}. The lattice parameters of h-BeH$_{\textit{n}}$ at 14 GPa are \textit{a} = 4.290(1) \AA, \textit{c} = 6.993(2) \AA, and \textit{V} = 111.5(1) \AA$^3$. (c) Raman scattering spectra of h-BeH$_{\textit{n}}$/H$_{2}$ mixture sample at 7.1 GPa, excess H$_{2}$, the Be after heated to 2000 K in Ar at 5 GPa, and the calculated spectrum for Be$_{4}$H$_{8}$(H$_{2}$)$_{2}$. The numbers indicate the correspondence between the observed and predicted peaks. The `dia.' is the signal from a diamond anvil.}
\end{figure*}

\begin{table*}[htb]
\caption{Structural parameters for Be$_{4}$H$_{8}$(H$_{2}$)$_{2}$ at 7.1 GPa. The lattice parameters and the Wyckoff position (Wyck. Pos.) of Be were determined from Rietveld refinement. The positions of H atoms are deduced by DFT calculation.}
\begin{ruledtabular}
 \begin{tabular}{lllllll}
       Structure & Atom & Wyck. Pos. & \textit{x}  &  \textit{y}  &  \textit{z}  &  Occ.\\
       \hline
    \textit{P}6$_3$/\textit{mmc}, \textit{Z} = 4 & Be   & 4\textit{f}  & 1/3 & 2/3 & 0.5610 & 1\\
    \textit{a} = 4.2965(2) \AA & H & 24\textit{l}  & 0.3796 & 0.4135 & 0.0522 & 1/4\\
     \textit{c} = 7.0095(6) \AA & H  & 12\textit{j}   & 0.5178 & 0.2894 & 1/4 & 1/6\\
    \textit{V} = 112.06 \AA$^{3}$ & H (mol.\footnotemark[1]) & 24\textit{l}   & 0.0258 & 0.0076 & 0.2582 & 1/12\\
                                                  & H (mol.\footnotemark[1])  & 24\textit{l}   & 0.1790 & 0.1837 & 0.3230 & 1/12\\
  \end{tabular}
  \end{ruledtabular}
            \footnotetext[1]{molecule}
\label{tb:lattice}
\end{table*}

\begin{figure}
 \includegraphics[scale=1]{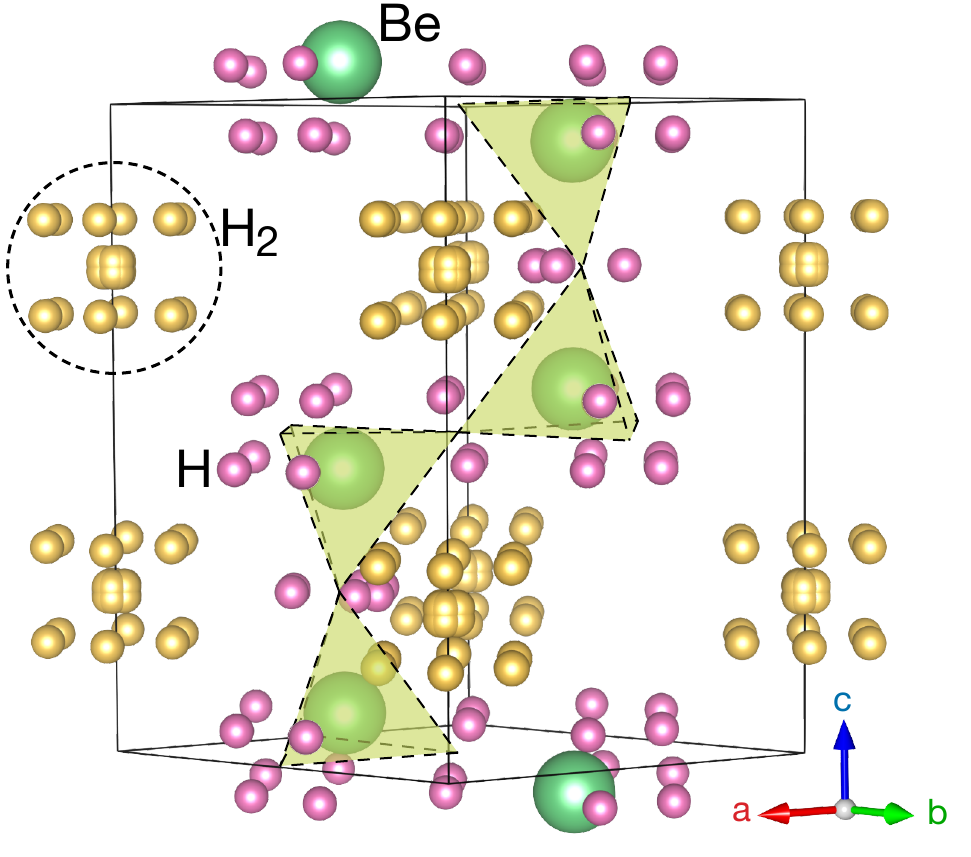}%
 \caption{\label{fig2}(color online) Be$_{4}$H$_{8}$(H$_{2}$)$_{2}$ (visualized using VESTA\cite{Momma:db5098}). Hydrogen atoms `H' are moving around the disordered positions. The H$_{2}$ are rotating in a dotted circle. Each tetrahedron is a corner-sharing BeH$_{4}$ unit. See the supplemental material for the Crystal structure Information File (CIF) of Be$_{4}$H$_{8}$(H$_{2}$)$_{2}$ \cite{supple}.}
 \end{figure}

We performed a structure analysis by indexing, space-group determination, and Rietveld refinement. 
The diffraction peaks were indexed using BIOVIA Materials Studio (MS) X-Cell software from Dassault Syst\`{e}mes \cite{Neumann}.
The crystal lattice and the atomic coordinates of Be were refined via Rietveld analysis using MS Reflex software \cite{BIOVIA}.
The energy stability and H-atom positions in each phase were investigated using the density functional theory (DFT) program MS CASTEP \cite{Clark2005}. 
We employed the generalized gradient approximation (GGA)-Perdew-Burke-Ernzerhof for solids (PBEsol) exchange-correlation functionals \cite{PhysRevLett.100.136406} and used ultrasoft pseudopotentials \cite{PhysRevB.41.7892}.
The lattice parameters were set to the experimental values and refined by Rietveld analysis, and the atomic positions were optimized to minimize the total energy. 
When we restricted the stoichiometry to BeH$_{2}$, we obtained two models; \textit{P}31\textit{c} and \textit{P}\={6}\textit{c}2 structures that are composed of corner-sharing four BeH$_{4}$ tetrahedrons.
However, the stress estimated by our DFT for these two models became 0 GPa in disagreement with experimental results (7.1 GPa). 
We note that these two models are isotypical to ice-Ih (ice at atmospheric pressure).
Besides, the \textit{V}$_{f.u.}$ of h-BeH$_{\textit{n}}$ is bigger than $\alpha$-BeH$_{2}$ by more than 7 \AA$^{3}$ at 7.1 GPa (Fig. 1b). 
Also,  h-BeH$_{\textit{n}}$ is less compressible than $\alpha$-BeH$_{2}$.
To mechanically support such a crystal structure with large internal space at pressures as high as 14 GPa, the presence of some substance in the interstitial sites is strongly suggested.
Considering that H$_{2}$ surrounds h-BeH$_{\textit{n}}$ in the sample chamber, H atom or H$_{2}$ molecule is the most probable.
When H atoms were at the interstitial sites of the $\textit{P}$31$\textit{c}$ and $\textit{P}$$\bar{6}$$\textit{c}$2, our DFT calculation and MD simulation calculated that BeH$_{4}$ tetrahedrons decomposed, and the unit cells collapsed.
When H$_{2}$ molecules were at the interstitial sites, an excellent fit to the experimental XRD profile was obtained, and the crystal structure became stable in the DFT calculation. 
The best structure model to describe h-BeH$_{\textit{n}}$ including the disorders of H and H$_{2}$ is summarized in the Table 1.
Note that the lattice parameters and atomic position of Be was determined from Rietveld refinement. 
The positions of H atoms are deduced by DFT calculation.
In Fig. 2, we show the visualized structure model. 
The BeH$_{4}$ tetrahedrons rotationally oscillate, and the unit cell goes back and forth between $\textit{P}$31$\textit{c}$ and $\textit{P}$$\bar{6}$$\textit{c}$2.
The H$_{2}$ molecules freely rotate.
See Movie 1 in the Supplemental Material for the rotational oscillations of BeH$_{4}$ units and H$_{2}$ molecules \cite{supple}. 

It is noted that we have concerned the possible occupation of the interstitial sites by Al and O from ruby, Re from the gasket, and carbon (C) from diamond anvil during the laser heating.
However, if such heavier atoms are included, the XRD measurements and the following analysis allow us to detect them.
Therefore, we exclude the occupation of the interstitial sites by C, O, and Al.

To examine the crystal structure model of h-BeH$_{\textit{n}}$, we compare the calculated Raman spectra for the suggested structure model with experimental results (Fig. 1c). 
We note that we employed a fully ordered structure with space group \textit{P}1 to do the calculation.
The peak $\sharp$16 consists of eight peaks originating from Be-H stretching modes  (See the supplemental Movie 2 \cite{supple} for the Be-H stretching mode at 2138 cm$^{-1}$.), and two of them at 2138 cm$^{-1}$ and 2157 cm$^{-1}$ are more intense than others.
We assume that the eight peaks overlap and form the observed single broad peak.
The peak  $\sharp$12 (1186 cm$^{-1}$) is the rotational mode of BeH$_{4}$ tetrahedron (See the supplemental Movie 3 \cite{supple} for the rotation that corresponds to $\sharp$12.).
The appearance of peaks $\sharp$12 and $\sharp$16 in the experiments strongly supports the suggested structure model. 
In the hydrides that consist of H$_{2}$, we sometimes observe the intra-H$_{2}$ stretching mode (vibron) at wavenumbers shifted from that of pure-H$_{2}$ (4200${-}$4400 cm$^{-1}$). 
In the present experiments, we did not observe the additional peaks near the vibron of excess-H$_{2}$.
We speculate that the peak from the accommodated  H$_{2}$ in h-BeH$_{\textit{n}}$ overlaps with that of the excess H$_{2}$. 
Although a perfect matching is not achieved, the relative intensities and relative positions of the observed and calculated peaks are in reasonable agreement as a whole. 
We conclude that h-BeH$_{\textit{n}}$ is Be$_{4}$H$_{8}$(H$_{2}$)$_{2}$ with the \textit{P}6$_{3}$/\textit{mmc} structure that is composed of BeH$_4$ tetrahedrons and H$_2$.

We investigated the structural stability of h-BeH$_{\textit{n}}$ on pressure decrease to 1 bar by monitoring the structure change using Raman scattering measurement (Fig. 3a and 3b).
We obtained the reproducible results from three independent experiments.
As pressure decreases, Raman scattering peaks show a monotonic decrease to 6.0 GPa (Fig. 3b). 
When the pressure reaches to 4.0 GPa, the peaks $\sharp$3 and $\sharp$11 disappear, and a peak appears at 432 cm$^{-1}$, and other peaks become broad.
However, the sample still maintains the peaks $\sharp$16 and $\sharp$12, suggesting that  BeH$_{4}$ tetrahedron persists at 4.0 GPa.
At 1 bar, the peaks from h-BeH$_{\textit{n}}$ vanish, and the peaks from Be metal and an unknown peak remain. 
Judging from the peak position, we consider that the possible origin of the unknown phase is the luminescence of ruby, covering the 100${-}$800 cm$^{-1}$ region,  that was there nearby the sample \cite{Syassen2008a}.
We conclude that h-BeH$_{\textit{n}}$ is unstable below 4 GPa decomposing to Be and H$_{2}$.

\begin{figure}
 \includegraphics[scale=1]{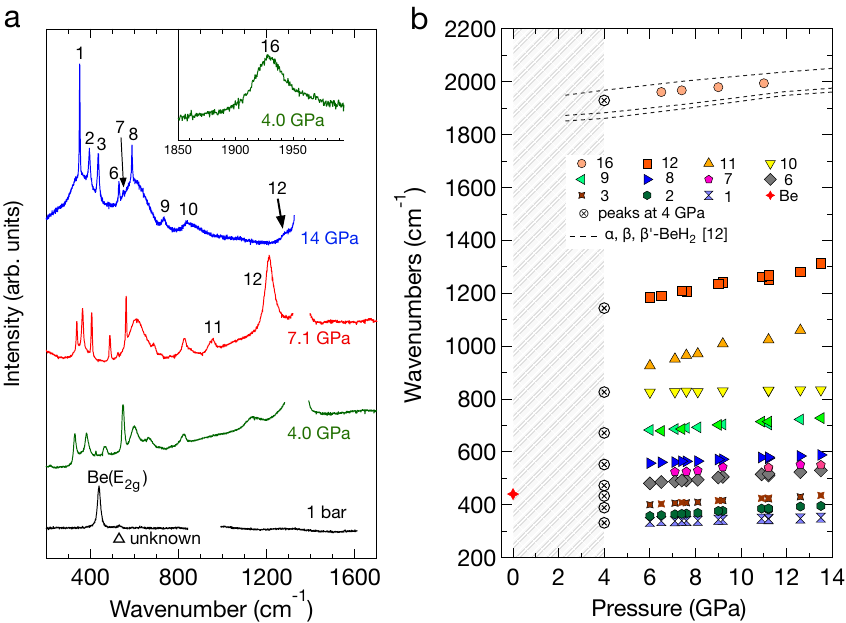}%
 \caption{\label{fig2}(color online) (a) The Raman scattering spectra of h-BeH$_{\textit{n}}$ as pressure decreases. The numbers indicate the peaks described in Fig. 2c. The spectra at 1 bar was measured after the diamond anvil was removed. (b) The pressure dependence of the Raman shifts for peaks $\sharp$1 ${-}$ $\sharp$16. The data obtained in three independent experiments are plotted together. Below 4 GPa (shaded area), h-BeH$_{\textit{n}}$ is unstable at room temperature. }
 \end{figure}

Here, we compare the crystal structure of h-BeH$_{\textit{n}}$ with $\alpha$-BeH$_{2}$ and discuss the Be-H bonding. 
Both of h-BeH$_{\textit{n}}$ and $\alpha$-BeH$_{2}$ consist of BeH$_{4}$ tetrahedrons.
The difference is in the positions of shared H atoms. 
In $\alpha$-BeH$_{2}$ at 1 bar, the H-Be-H connection is bent about 130 deg., and a Be-H distance is 1.38 {\AA} while another is 1.41 {\AA} \cite{Pepin2016}.
In h-BeH$_{\textit{n}}$, the Be-H-Be angle is near 180 deg.
The H atoms are at the midpoint between two Be atoms, and the Be-H distance (1.44 {\AA} at 7 GPa) is longer than that of $\alpha$-BeH$_{2}$.
Another difference is in the structural stability at 1 bar.
If h-BeH$_{\textit{n}}$ keeps covalence in the Be-H bond, it would transform into $\alpha$-BeH$_{2}$ or amorphous BeH$_{2}$ at pressures below 4 GPa.
Thus, we think that the shared H atom is not tightly bonded to a single Be atom, and the Be-H bond in h-BeH$_{\textit{n}}$ is not covalent.
We speculate that the accommodated H$_{2}$ molecules mechanically support the unit cell of h-BeH$_{\textit{n}}$ and also elongate the Be-H bond to the extent the bond loses covalency.

To go into the interaction between H$_2$ and surrounding BeH$_4$ tetrahedrons, we also compare the crystal structure of h-BeH$_{\textit{n}}$ with the compound of water ice and helium (He hydrate).
Teeratchanan and Hermann have performed the first principle study assuming van der Walls interaction between He and H$_2$O with binding energy \(-7.99\) meV and predicted the formation of He hydrate with ice-Ih structure where He occupies the interstitial site (cage) \cite{Teeratchanan2015}. 
From the similarity of crystal structure and the free rotation of H$_2$, we speculate that the interaction between H$_2$ and BeH$_2$ is similar to that of the He-hydrate.

To obtain further knowledge about the thermodynamical stability of Be$_{4}$H$_{8}$(H$_{2}$)$_{2}$, we searched for the most stable crystal structure at static conditions using a genetic algorithm and first-principles calculations (See the Supplementary Material \cite{supple}).
Our structure search predicts that BeH$_3$ has a positive formation enthalpy of 12 mRy/atom at 20 GPa for decomposition into BeH$_2$ and H$_2$, as is the case in the theoretical results reported earlier \cite{Vajeeston2004, Wang2014, Yu2014, Hooper2013}.
Besides, a triclinic \textit{P}$\bar{1}$ structure emerges as the most stable one instead of \textit{P}6$_3$/\textit{mmc} observed by experiments.
More than 80\% of the compounds included in the Inorganic Crystal Structure Database have positive formation energy less than 36 meV/atom \cite{Wu2013a}.
The formation enthalpy of BeH$_3$ is below the threshold for the acceptable instability, which suggests that Be$_{4}$H$_{8}$(H$_{2}$)$_{2}$ with \textit{P}6$_3$/\textit{mmc} is a metastable phase that is stabilized dynamically by the thermal effect at finite temperature.
This would be the reason for the experimental result that Be$_{4}$H$_{8}$(H$_{2}$) is less dense than $\alpha$-BeH$_{2}$+\((1/2)\)H$_2$ (Fig. 1b).


In summary, Be$_{4}$H$_{8}$(H$_{2}$)$_{2}$ is formed by applying high pressure (5${-}$8 GPa) and high temperature (above 1700 K) to the Be/H$_{2}$ mixture.
In Be$_{4}$H$_{8}$(H$_{2}$)$_{2}$, BeH$_{4}$ tetrahedrons form a hexagonal \textit{P}6$_{3}$/\textit{mmc} structure, and there are H$_{2}$ molecules in the interstitial sites. 
The observed Be$_{4}$H$_{8}$(H$_{2}$)$_{2}$ is possibly a metastable phase in the Be-H phase diagram.
The H atoms are not covalently bonded to Be atoms.
The present results would motivate further studies such as searching for superconductivity and recovering Be$_{4}$H$_{8}$(H$_{2}$)$_{2}$ to ambient pressure.
So far, there has been no prediction of the formation of Be$_{4}$H$_{8}$(H$_{2}$)$_{2}$.
Revealing the phase diagram of the Be-H system would be a particular interest to understand the interaction between Be and H.

\begin{acknowledgments}
This study was supported by JSPS KAKENHI Young Scientists (B) (25800195),
Scientific Research (C) (17K05541), and Specially Promoted Research (26000006),
and MEXT under ESICMM (12016013) and `Exploratory Challenge on Post-K computer' (Frontiers of Basic Science: Challenging the Limits). 
Authors thank for the awarded beam time for XRD measurements at BL10XU/SPring-8 (Proposal Nos. 2013B1109, 2014A1320).
\end{acknowledgments}

\providecommand{\noopsort}[1]{}\providecommand{\singleletter}[1]{#1}%

\end{document}